\documentclass{webofc}
\usepackage[varg]{txfonts}  
\usepackage{hyperref}
\usepackage{paralist} 
\usepackage{lineno}

\begin{document}
	
\title{Jiskefet, a bookkeeping application for ALICE}
\author{
	\firstname{Marten} \lastname{Teitsma}\inst{1}\email{m.teitsma@hva.nl}
	\and
	\firstname{Vasco Chibante} \lastname{Barosso}\inst{2}
	\and
	\firstname{Pascal} \lastname{Boeschoten}\inst{1}
	\and
	\firstname{Patrick} \lastname{Hendriks}\inst{1}
}
\institute{Amsterdam University of Applied Sciences
	\and
	CERN 
}

\abstract{A new bookkeeping system called Jiskefet is being developed for A Large Ion Collider Experiment (ALICE) during Long Shutdown 2, to be in production until the end of LHC Run 4 (2029). Jiskefet unifies two functionalities: \begin{inparaenum}[a)] \item gathering, storing and presenting metadata associated with the operations of the ALICE experiment and \item tracking the asynchronous processing of the physics data. \end{inparaenum} It will replace the existing ALICE Electronic Logbook and AliMonitor, allowing for a technology refresh and the inclusion of new features based on the experience collected during Run 1 and Run 2. The front end leverages web technologies much in use nowadays such as TypeScript and NodeJS and is adaptive to various clients such as tablets, mobile devices and other screens. The back end includes an OpenAPI specification based REST API and a relational database. This paper will describe the organization of the work done by various student teams who work on Jiskefet in sequential and parallel semesters and how continuity is guaranteed by using guidelines on coding, documentation and development. It will also describe the current status of the development, the initial experience in detector stand-alone commissioning setups and the future plans.
}
\maketitle

\section{Introduction}
During the LHC Long Shutdown 2 a renewal of the bookkeeping systems in place for ALICE is envisioned. For this, a unified experience for shifters active in detector operations, physicists searching the run catalogue to find appropriate runs and overall management of the ALICE operation should be provided. Users access all metadata related to operational activities, keep historical records of configurations used to operate the detector, execute performance studies of detector operations, monitor the quality of the collected data, and produce reports for operational teams and experiment management in a single place.

ALICE \cite{aamodt2008alice} is one of the four main detectors collecting collision data at the CERN LHC. ALICE focuses on the study of quark--gluon plasma in heavy-ion collisions but also carries out substantial measurements of proton--proton interactions. ALICE produces a lot of data of interest to physicists who need to know the circumstances under which the data are gathered. For this, metadata is produced and stored by the bookkeeping system. Secondly, the goal of the bookkeeping system is to keep track of the data stored on the Grid and used for analysis by end-users. 

At the start of the ALICE operations in 2007 a bookkeeping system was developed and it evolved with time during the LHC data-taking periods Run 1 (2009-2013) and Run 2 (2015-2018) following the users needs. This system was based on the LAMP (Linux, Apache, MySQL and PHP) software stack with a relational database, a web-based Graphical User Interface for the members of the ALICE collaboration, and bindings for machine to machine access \cite{altini2010alice}. The electronic logbook system thus far gathered informations from about 5,000 LHC fills, 280,000 data-taking runs, and comprises about 37 GB data from 195,000 log entries and 20,000 file attachments. For monitoring the data on the Grid a system called AliEn, later known as AliMonitor, was developed to present detailed information such as CPU and memory usage, open files and network traffic of each site offering services \cite{bagnasco2008alien}. The Data Preparation Group, Physics Working Groups and ALICE physicists are the main users of AliMonitor. Since its deployment it has stored 12.9 million jobs, i.e. specified activities on data, amounting to 97 GB data.

In this paper we first present the methods for developing the application and the tools we used. Then the requirements of the system are described and the architecture of the system is shown. We finish with the current status, discussion, future work and conclusions.

\section{Methods and tools}
To gather the requirements for the new bookkeeping system we started with a series of interviews at CERN. The requirements are described in Section \ref{sec:Requirements}. As a method of development we used Scrum \cite{scrum2020}. 

The development of Jiskefet is performed by several student teams. The main effort is done in Amsterdam at the Amsterdam University of Applied Sciences (AUAS). Several teams and interns involved in the Software for Science initiative work on the application consecutively. To work effectively it is important to transfer knowledge and skills in a transparent way. Documentation is very important together with the continuous involvement of staff. After the first release, student teams from a course of Web Development and the Moscow Polytechnic University were involved. The first team worked on a design for the GUI. The team from Moscow worked on the reporting functionality.

To organise the communication among the several teams we used regular video conferences, Slack, Telegram and the odd email. Visits to the different Institutes contributing to the project were also organised.

Git is used for software version control. The hosting of the code is done on GitHub \cite{jiskefet2020} as well as a private repository on a GitLab instance. The formal documentation is  edited in LaTeX. Less formal documentation is included in the repository as Markdown files. Coordination of the project is performed with the Jira online tool \cite{jira2020}.

Testing is part of the continuous integration process for which we use Travis CI \cite{travis2020} and Jenkins \cite{jenkins2020}. The testing itself is done by Mocha \cite{mocha2020} and Jest \cite{jest2020}. The linting is done by TSLint \cite{tslint2020}. We use CodeCov \cite{codecov2020} to determine how much of the code is being tested.
	
\section{Requirements}
\label{sec:Requirements}
To retrieve the requirements we interviewed several stakeholders of the system at the start of the project. We then categorised the findings in a System Requirements Specification according to IEEE Std 830-1998 \cite{ieeeReq1998}. In this specification the business goals, vision and scope are described. Several interfaces, functions and users are mentioned. Furthermore, the functional and non-functional requirements are summed up.

The functional requirements are categorised according to the large number of various users of the system. To name just a few: shifter, (sub system) run coordinator, manager, member of the physics board. Each user has its own specific requirements. Besides the existing functionalities of the electronic logbook system in place, several new requirements were mentioned during the interviews such as: users should be able to make a log entry using a template to avoid incomparability, it should be possible to have a flexible tagging system for runs, a smart editor should be available, a search functionality should be in place. 

The bookkeeping system shall function as part of a large information technology system. Several system attributes, which should lead to a friction-less embedding, were mentioned during the requirement engineering phase. The choice of tools and the software stack was dependent on several desiderata such as:
\begin{itemize}
	\item familiarity with the software at CERN;
	\item minimisation of dependencies;
	\item maximisation of development community.
\end{itemize}
It was also important to use modern tools and software expected to remain in use for the entire time of the project.

Jiskefet has to be in production until 2029. Since during this period several students and members of the staff will independently be involved in the development and maintenance, sustainability of the application is of importance. An application can be characterised by its intrinsic sustainability, i.e. level of documentation, testing, readability, usage of third party libraries, usefulness and scalability. Its extrinsic sustainability is characterised by its availability, resourcefulness, level of community actions and relations, independence from infrastructure \cite{deSouza2014}.

The availability of the system should be optimal and not be a reason for the detector to stop taking data. Each entry,  i.e. input of data, should be logged and although the GUI and REST API are not essential the database should be highly available. Once in the repository, no data should be lost.

\section{Architecture}
\begin{figure}[ht]
	\includegraphics[scale=0.3]{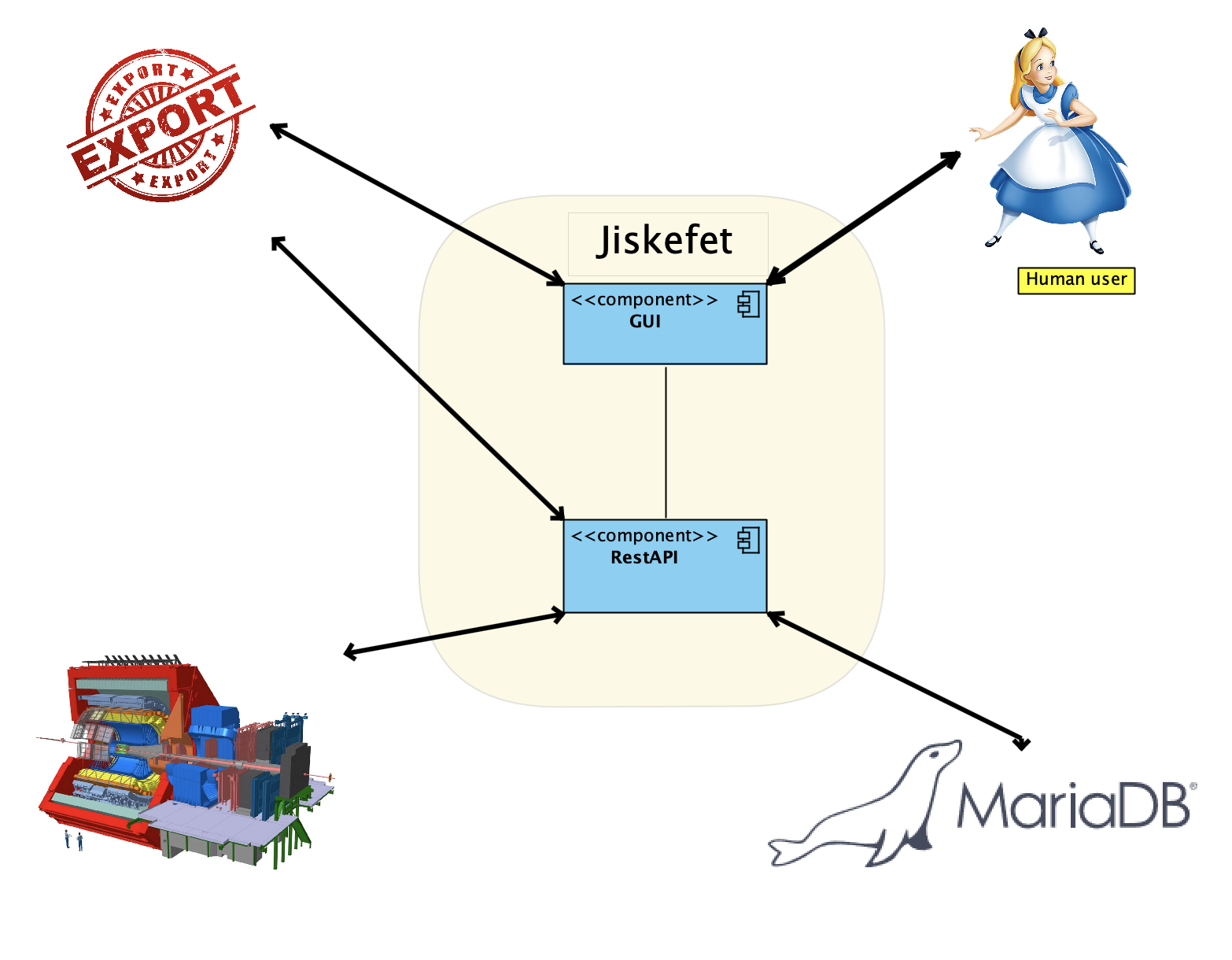}
	\caption{Environment of Jiskefet}
	\label{fig:Environment of Jiskefet}
\end{figure}

The bookkeeping system used in ALICE during Run 1 and 2 used a web technology and the same requirement is valid for the new system. The web has evolved fast and so did the technology behind it. Users now have more seamless and responsive applications in the browser with the rising popularity of Single Page Application (SPA) frameworks and real time web apps developed in JavaScript. These techniques require strict separation of data and markup. The data is delivered and gathered by a RESTful back-end as can be seen in Figure \ref{fig:Environment of Jiskefet}. The environment of Jiskefet consists of a database, the detector, human users who directly interact with the system and users who interact not directly but use other applications to retrieve data or send requests.

The JavaScript technology stack is notorious for its many dependencies. To mitigate the number of dependencies, we carefully assessed common frameworks and libraries for developing Jiskefet. These frameworks where selected based on a popularity ranking and on the features they packed. The ALICE team was working on a collection of frameworks with a CERN wrapper around it called WebUI \cite{webui2020}. Whilst WebUI is merely a collection of relatively lightweight nodeJS \cite{nodejs2020} modules i.e. express.js \cite{expressjs2020}, mySql, the volatility of WebUI at that time made us choose for more popular frameworks, i.e. Nest.js \cite{nestjs2020} and Mithril \cite{mithril2020}.

Mithril is a modern client-side Javascript framework for building Single Page Applications. It's small, fast and provides routing and several utilities out of the box. The benefit of choosing Mithril over for example React is that the ALICE people are familiar with it because it is included in WebUI.

One of the main non-functional requirements is to develop a REST-API as a back-end. RESTful Web services are constrained to be a client-server architecture, stateless, cacheable, layered and have a uniform interface. Advantages are good performance, scalability, simplicity, modifiability, etc \cite{fielding2000architectural}. The back-end is implemented with the Nest.js web framework. One of the main advantages is the Swagger \cite{swagger2020} functionality built in. Swagger helps us and other developers to model, document and test the API.

In the data model several entities are discerned. Entities concerning the work done at ALICE are an `Activity', i.e. a set of well-defined tasks during a finite time period. An example is a `Run'. An `Activity' aggregates configurations, selected tasks, global statistics etc. A `Run' can be of different types such as `Detector Calibration Run' or `Global Run'. A `Run' is formally defined as `a synchronous data readout and processing Activity in the $O^2$ farm with a specific and well-defined configuration'. It normally ranges from a few minutes to tens of hours and provides a unique identifier for the data set generated at the end of the synchronous data flow. Another `Activity' is a `Reconstruction Pass' which is defined as `asynchronous data processing Activity in the $O^2$ farm or the Grid with a specific and well-defined configuration'. Runs identify the data generated at the end of the synchronous data flow. Reconstruction Passes will have as input the data generated in either a Run or a previous Reconstruction Pass. Runs can be executed either during an LHC Fill (as for most global physics data taking cases) or in between LHC Fills (for cosmics data taking, calibration and/or tests). An LHC Fill can have many Runs but a Run only belongs to an LHC Fill. Reconstruction Passes are indirectly connected to LHC Fills via the Run. 

The database is implemented using MariaDB \cite{mariadb2020}. Familiarity with its predecessor MySql and ease of use were the main reasons for this choice. For the users the central entity is the log entry. Log Entries are generic and can be connected to any of the other entities. 

\begin{figure}[ht]
	\includegraphics[scale=0.3]{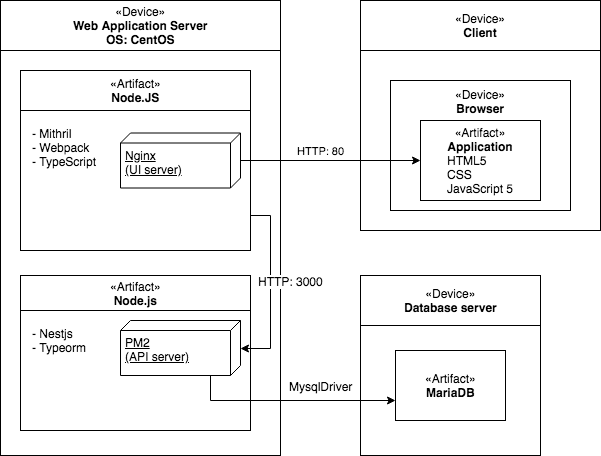}
	\caption{Deployment diagram}
	\label{fig:deployment_diagram}
\end{figure}
Jiskefet consists of several packages as shown in Figure \ref{fig:deployment_diagram}. On a CentOS platform a Nginx Web Server \cite{nginx2020}, consisting of two Node.JS applications, is deployed. For the User Interface we use Mithril as a framework, Webpack \cite{webpack2020} to bundle the modules and decrease the dependencies and TypeScript \cite{typescript2020} as a way to make JavaScript more consistent. On the front-end we use HTML5, CSS and JavaScript. For the back-end we use NestJs and Typeorm \cite{typeorm2020} as an interface with the database. For loadbalancing PM2 \cite{pm22020} is used. As a database we use MariaDB. To streamline the work flow and ease testability we use Ansible \cite{ansible2020} as automated deployment framework. 

\section{Current Status and Discussion}
A prototype has been developed. The prototype was to be distributed among the teams performing detector commissioning of the ALICE subdetectors, but it wasn't because of a number of mismatches in the Ansible playbook. In web development the use of external libraries is normal practice. This creates an important drawback making it obligatory to do maintenance continuously. Particularly NodeJS has a great number of dependencies. Together with other packages it makes the current version too much dependent on insecure sources. Although thoroughly selected packages, there is always the risk of depending on outdated or obsolete packages in the long run. A plan is needed to cover those risks. 

Communication between the several teams and stakeholders in this project is of great importance. Insufficient feedback on what has been produced, shifting requirements or misunderstandings, among other problems make a project of this size and timespan complex. 

Important decisions about the use of specific tools were made when at ALICE the development of a web framework was not yet matured. The asynchronous development and use of different frameworks  
is causing now some disturbance in the working together of the systems. This and other problems make it seemingly unavoidable that the current prototype has to be refactored. 

\section{Future Developments }
To comply with the non-functional requirements of the user we will refactor Jiskefet to include WebUI. Furthermore, we are now going to work on the second goal which is the management of the computation of the data on the Grid.

\section{Conclusions}
The first goal of our project was to create a log functionality for ALICE. We now have a first version which has to be refactored. The development of such a fairly complex system for a remote client by several student teams from different countries is not a trivial project. A lot of time has to be spent on communication to make sure everybody involved knows of what the projectpartner is talking about. 

\bibliography{alice}

\begin{thebibliography}{28}

\bibitem{aamodt2008alice}
K.~Aamodt, A.A. Quintana, R.~Achenbach, S.~Acounis, D.~Adamov{\'a}, C.~Adler,
  M.~Aggarwal, F.~Agnese, G.A. Rinella, Z.~Ahammed et~al., \emph{The ALICE
  experiment at the CERN LHC} (IOP Publishing, 2008), Vol.~3, p. S08002

\bibitem{altini2010alice}
V.~Altini, F.~Carena, W.~Carena, S.~Chapeland, V.C. Barroso, F.~Costa,
  R.~Divi{\`a}, U.~Fuchs, I.~Makhlyueva, F.~Roukoutakis et~al., \emph{The ALICE
  electronic logbook}, in \emph{Journal of Physics: Conference Series} (IOP
  Publishing, 2010), Vol. 219, p. 022027

\bibitem{bagnasco2008alien}
S.~Bagnasco, L.~Betev, P.~Buncic, F.~Carminati, C.~Cirstoiu, C.~Grigoras,
  A.~Hayrapetyan, A.~Harutyunyan, A.~Peters, P.~Saiz, \emph{AliEn: ALICE
  environment on the GRID}, in \emph{Journal of Physics: Conference Series}
  (IOP Publishing, 2008), Vol. 119, p. 062012

\bibitem{scrum2020}
\emph{Scrum alliance},
  \url{https://www.scrumalliance.org/about-scrum/overview}, accessed: March
  2020

\bibitem{jiskefet2020}
\emph{Jiskefet}, \url{https://github.com/SoftwareForScience}, accessed: March
  2020

\bibitem{jira2020}
\emph{Jira software}, \url{https://www.atlassian.com/software/jira}, accessed:
  March 2020

\bibitem{travis2020}
\emph{Travis ci}, \url{https://travis-ci.org}, accessed: March 2020

\bibitem{jenkins2020}
\emph{Jenkins}, \url{https://jenkins.io}, accessed: March 2020

\bibitem{mocha2020}
\emph{Mocha}, \url{https://mochajs.org}, accessed: March 2020

\bibitem{jest2020}
\emph{Jest}, \url{https://jestjs.io}, accessed: March 2020

\bibitem{tslint2020}
\emph{Tslint}, \url{https://github.com/palantir/tslint}, accessed: March 2020

\bibitem{codecov2020}
\emph{Codecov}, \url{https://codecov.io}, accessed: March 2020

\bibitem{ieeeReq1998}
IEEE, \emph{830-1998 - ieee recommended practice for software requirements
  specifications} (2009),
  \urlstyle{tt}\url{"https://standards.ieee.org/standard/830-1998.html"}

\bibitem{deSouza2014}
M.R. de~Souza, R.~Haines, C.~Jay, \emph{Defining sustainability through
  developers’ eyes: Recommendations from an interview study}, in
  \emph{Proceedings of the 2nd Workshop on Sustainable Software for Science:
  Practice and Experiences (WSSSPE 2014)} (2014)

\bibitem{webui2020}
\emph{Webui}, \url{https://github.com/AliceO2Group/WebUi}, accessed: March 2020

\bibitem{nodejs2020}
\emph{Node.js}, \url{https://nodejs.org/en/}, accessed: March 2020

\bibitem{expressjs2020}
\emph{Express.js}, \url{https://expressjs.com}, accessed: March 2020

\bibitem{nestjs2020}
\emph{Nest.js}, \url{https://nestjs.com}, accessed: March 2020

\bibitem{mithril2020}
\emph{Mithril}, \url{https://mithril.js.org}, accessed: March 2020

\bibitem{fielding2000architectural}
R.T. Fielding, R.N. Taylor, \emph{Architectural styles and the design of
  network-based software architectures}, Vol.~7 (University of California,
  Irvine Irvine, USA, 2000)

\bibitem{swagger2020}
\emph{Swagger}, \url{https://swagger.io}, accessed: March 2020

\bibitem{nginx2020}
\emph{Nginx}, \url{https://nginx.org}, accessed: March 2020

\bibitem{webpack2020}
\emph{Webpack}, \url{https://webpack.js.org}, accessed: March 2020

\bibitem{typescript2020}
\emph{Typescript}, \url{https://www.typescriptlang.org}, accessed: March 2020

\bibitem{typeorm2020}
\emph{Typeorm}, \url{https://opencollective.com/typeorm}, accessed: March 2020

\bibitem{pm22020}
\emph{Pm2}, \url{https://www.npmjs.com/package/pm2}, accessed: March 2020

\bibitem{mariadb2020}
\emph{Mariadb}, \url{https://mariadb.org}, accessed: March 2020

\bibitem{ansible2020}
\emph{Ansible}, \url{https://www.ansible.com}, accessed: March 2020

\end{thebibliography}

\end{document}